%% file: main.tex
\documentclass[sigconf,nonacm]{acmart}

\usepackage{booktabs} 
\usepackage{paralist}

\usepackage{mathptmx} 

\usepackage{listings}
\input ./style/pkgs
\input ./style/symbols

\input ./style/spacing

\usepackage[capitalise]{cleveref}

\usepackage{flushend}
\usepackage{balance}

\usepackage{tabularx}
\usepackage[labelfont=bf]{caption}

 \makeatletter
\AfterEndEnvironment{algorithm}{\let\@algcomment\relax}
\AtEndEnvironment{algorithm}{\kern2pt\hrule\relax\vskip3pt\@algcomment}
\let\@algcomment\relax
\newcommand\algcomment[1]{\def\@algcomment{\footnotesize#1}}

\renewcommand\fs@ruled{\def\@fs@cfont{\bfseries}\let\@fs@capt\floatc@ruled
  \def\@fs@pre{\hrule height.8pt depth0pt \kern2pt}%
  \def\@fs@post{}%
  \def\@fs@mid{\kern2pt\hrule\kern2pt}%
  \let\@fs@iftopcapt\iftrue}
\makeatother

\begin{document}
\title{Spatter: A Tool for Evaluating Gather / Scatter Performance}

\author{Patrick Lavin$^\dagger$, Jeffrey Young$^\dagger$, Jason Riedy$^\dagger$, Rich Vuduc$^\dagger$}
\email{{plavin3,jyoung9,jason.riedy,richie}@gatech.edu}

\author{Aaron Vose$^\ddagger$, Dan Ernst$^\ddagger$}
\email{avose@aaronvose.net, daniel.ernst@hpe.com}

\affiliation{%
\institution{Georgia Institute of Technology$^\dagger$ \hspace{0.3em} Cray, Inc.$^\ddagger$}
}



\newcommand{\gatherscatter}{gather / scatter\xspace}
\newcommand{\gatherscatterns}{gather / scatter\xspace}
\newcommand{\GatherScatter}{Gather / Scatter\xspace}
\newcommand{\GS}{G/S\xspace}

\input{sections/outline.tex}
\newpage

\maketitle

\thispagestyle{plain}
\pagestyle{plain}

\input{sections/abstract.tex}
\input{sections/intro.tex}
\input{sections/app_characterization.tex}
\input{sections/benchmark_design.tex}

\input{sections/exp_setup.tex}
\input{sections/results.tex}

\input{sections/related-work.tex}
\balance
\input{sections/future_work.tex}

\input{sections/conclusion.tex}
%


\newpage
\bibliographystyle{ACM-Reference-Format}
\bibliography{bib/refs} 

\newpage
\onecolumn
\appendix
\input{sections/appendix}



\end{document}

%% file: style/pkgs.tex
\usepackage{style/my-stdpkgs}
\usepackage{style/my-typography}
\usepackage{acronym}
\usepackage{microtype} 
\usepackage{afterpage}

\usepackage[T1]{fontenc}
\usepackage[latin9]{inputenc}
\usepackage{verbatim}
\usepackage{textcomp}

\usepackage{array,ragged2e}

\usepackage{float}
\usepackage{amsfonts}
\usepackage{morefloats}

\makeatletter
\def\@IEEEsectpunct{.\ \,}
\def\paragraph{\@startsection{paragraph}{4}{\z@}{1.5ex plus 1.5ex minus 0.5ex}%
{0ex}{\normalfont\normalsize\sffamily\bfseries}}
\makeatother

\usepackage{algorithm}
\usepackage[noend]{algorithmic}

\acrodef{PCIe}{Peripheral Component Interface Express}



%% file: style/symbols.tex
\acrodef{DVFS}[DVFS]{dynamic frequency and voltage scaling}

\acrodef{DFS}[DFS]{dynamic frequency scaling}

\acrodef{SSSP}[SSSP]{single-source shortest path}

\acrodef{GPU}[GPU]{graphics co-processors}

%% file: style/spacing.tex
\renewenvironment{thebibliography}[1]{%
   \begin{oldthebibliography}{#1}%
     \setlength{\itemsep}{-.3ex}%
}%
{%
   \end{oldthebibliography}%
}



%% file: sections/outline.tex





%% file: sections/abstract.tex
This paper describes a new benchmark tool, Spatter, for
assessing memory system architectures in the context of
a specific category of indexed accesses known as gather 
and scatter. These types of operations are 
increasingly used to express sparse and irregular 
data access patterns, and they have widespread utility in many modern HPC
applications including scientific simulations, data
mining and analysis computations, and graph processing. However, 
many traditional benchmarking tools like STREAM, STRIDE, and GUPS 
focus on characterizing only uniform stride or fully random accesses despite evidence that
modern applications use varied sets of more complex access patterns.  

Spatter is an open-source benchmark that provides a tunable and configurable framework to benchmark a variety of indexed access patterns, including variations of \gatherscatter that are seen in HPC mini-apps evaluated in this work. The design of Spatter includes tunable backends for OpenMP and CUDA, and experiments show how it can be used to evaluate 1) uniform access patterns for CPU and GPU, 2) prefetching regimes for gather/scatter, 3) compiler implementations of vectorization for gather/scatter, and 4) trace-driven ``proxy patterns'' that reflect the patterns found in multiple applications. The results from Spatter experiments show that GPUs typically outperform CPUs for these operations, and that Spatter can better represent the performance of some cache-dependent mini-apps than traditional STREAM bandwidth measurements.

%% file: sections/intro.tex
\section{Introduction} \label{sec:intro}
We consider the problem of how to assess the performance of modern memory systems with respect to \emph{indexed memory accesses}, such as gather and scatter (\GS) operations.
Our motivation derives from both applications and hardware.
On the application side, there are many instances where memory operations involve loads or stores through a level of indirection (e.g., \texttt{reg $\leftarrow$ base[idx[k]]}).
For instance, such indexed memory access is common in scientific and data analysis applications that rely on sparse and adaptive data abstractions, including adaptive meshes, sparse matrices and tensors, and graphs, which are our focus.
On the hardware side, new CPU architectures have begun to incorporate advanced vector functionality
like AVX-512 and the Scalable Vector Extension (SVE) for improving SIMD application performance.

Within this context, our strategy to understanding the interactions between application-relevant \GS operations and modern hardware relies on the development of a microbenchmarking tool.
It aims to express critical features of real \GS workloads, derived from applications but abstracted in a way that is easy to adopt by system-oriented stakeholders. 
These include
\begin{inparaenum} 
  \item vendors and hardware architects might wonder how new ISAs (such as AVX-512) and their implementation choices actually impact memory system performance;
  \item application developers may consider how the data structures they choose impact the \GS instructions their code compiles to; and
  \item compiler writers might require better data on real-world memory access patterns to decide whether to implement a specific vectorization optimization for sparse accesses.
\end{inparaenum}
Although these groups could turn to any number of memory-focused microbenchmarks available today~\cite{mcalpin:2018:streamingstore}, we believe a gap still exists in the focused evaluation of system performance for indexed accesses, including \gatherscatter workloads.



In light of these needs, we have formulated a new microbenchmarking tool called Spatter.
It evaluates indexed access patterns based on \GS operations informed by applications across different language and architecture platforms.
More importantly, we believe Spatter can help to answer a variety of system, application, and tool evaluation questions, some of which include:
\begin{inparaenum}
\item What application \GS patterns exist in the real world, and how do they impact memory system performance?
\item How does prefetching affect performance of indexed accesses on modern CPU platforms?
\item How does the performance of \GS change when dealing with sparse data on CPUs and GPUs?
\end{inparaenum}

The design of the Spatter tool suite aims to address these questions.
At a basic level, Spatter provides \textbf{tunable gather and scatter implementations}.
These include \textbf{CUDA and OpenMP backends} with knobs for adjusting thread block size and ILP on GPUs and work-per-thread on CPUs.
Spatter also includes a \textbf{scalar, non-vectorized backend} that can serve as a baseline for evaluating the benefits of vector load instructions over their scalar counterparts. 
Lastly, Spatter has built-in support for running parameterized memory access patterns and custom patterns.
We show, for instance, how one can collect \GS traces from Department of Energy (DoE) mini-apps to gain insights or make rough predictions about performance for hot kernels that depend on indexed accesses (\cref{sec:app_char}).

Our initial experimental evaluations of Spatter show that newer GPU architectures perform best for both gather and scatter operations in part due to memory coalescing and faster memories.
AMD Naples performs best of all the CPU-based platforms (Broadwell, Skylake, TX2) for strided accesses.
A study of prefetching with Spatter further shows how \gatherscatter benefits from modern prefetching across Broadwell and Skylake CPUs.
Spatter's scalar backend is also used to demonstrate how compiler vectorization can improve \GS with large improvements for both Skylake and Knight's Landing (\cref{ssec:scalar_simd}).
Experiments for three DoE mini-apps show \GS performance improvements enabled by caching on CPU systems and by fast HBM memory on GPUs.
These parameterized access pattern studies also suggest that STREAM bandwidth does not correlate well with specific mini-apps that are cache-dependent, which further motivates benchmarks like Spatter that do better.

\begin{table*}[ht]
\centering
\caption{High-Level Characterization of Application \GS Patterns.}
\label{table:miniapp-char}
\begin{small}
\begin{tabular}{lrrr}
\toprule
   \textbf{Application} (Extracted Patterns) \texttt{Selected Kernels} $\:\:$ & Gathers & Scatters & \GS MB (\%) \\
\midrule
  \textbf{AMG} (mostly stride-1) & & & \\
  $\:\:\:$\small{\texttt{hypre\_CSRMatrixMatvecOutOfPlace}} & 1,696,875 & 0 & 217 (17.8) \\
  \textbf{LULESH} (uniform-stride) & & & \\
  $\:\:\:$\small{\texttt{IntegrateStressForElems}} & 828,168 & 382,656 & 155 (22.4) \\
  $\:\:\:$\small{\texttt{InitStressTermsForElems}} & 1,121,844 & 1,153,827 & 291 (67.6) \\
  \textbf{Nekbone} (uniform-stride) & & & \\
  $\:\:\:$\small{\texttt{ax\_e}} & 2,948,940 & 0 & 377 (33.3) \\
  \textbf{PENNANT} (fixed-stride, broadcast)  & & & \\
  $\:\:\:$\small{\texttt{Hydro::doCycle}} & 728,814 & 0 & 93 (13.9) \\
  $\:\:\:$\small{\texttt{Mesh::calcSurfVecs}} & 324,064 & 0 & 41 (39.5) \\
  $\:\:\:$\small{\texttt{QCS::setForce}} & 891,066 & 0 & 114 (45.5) \\
  $\:\:\:$\small{\texttt{QCS::setQCnForce}} & 1,214,318 & 323,800 & 197 (64.5) \\
\bottomrule
\end{tabular}
\end{small}
\end{table*}

%% file: sections/app_characterization.tex
\section{\GatherScatter in Real-World Applications} \label{sec:app_char}

To motivate our interest in \gatherscatter performance, we studied several prominent DOE mini-apps from the CORAL and CORAL-2 procurements~\cite{coral,coral2}.
Such software provides a rich source of information about the computational and memory behavior requirements of critical scientific workloads in both government as well as academic environments.
Many of these workloads contain important kernels which stress \gatherscatter performance.
Indeed, one aim of Spatter is to leverage such mini-apps as a source of real-world \GS patterns.

\Cref{table:patterns} in the Appendix provides detailed information on the specific patterns extracted from these applications, so we focus on the high-level characteristics of each application in this section. We note that many of these patterns are \textit{complex} in either their delta offset between \gatherscatter or their type (broadcast, stride-N, mostly stride-1) as is discussed more in \Cref{sec:bmk_design} and \Cref{sec:results}.

%
%
\IGNORE{Workloads prevalent in the field of high-performance computing frequently depend heavily upon \gatherscatter performance compared to those encountered elsewhere. Additionally, such workloads increasingly encounter energy or power consumption as a bottleneck \cite{kambatla2014trends}.
Consider that data movement dominates energy consumption; a 7nm process requires around 50 pJ to load, compute, and write, while the computation alone only requires approximately 17 pJ \cite{borkar2013role}. Data motion is particularly expensive in the context of ``dark bandwidth'' \cite{beard2017eliminating}, in which data is moved using a larger granularity than that required by the final computation. 
Often, an entire 64 byte cache line (or more) is moved even if a single byte in the line is actually used. Indeed, cache line utilization is low for many applications \cite{srinivasan2011improving}.

The U.S. Department of Energy (DoE) releases HPC mini-apps and benchmarks as part of the acquisition process for some of the largest world-class supercomputing systems, such as those acquired through the CORAL \cite{coral} and CORAL-2 \cite{coral2} procurements.
Such software is a rich source of information w.r.t. the computational requirements of critical scientific workloads in both governmental as well as academic environments.
Many of these workloads contain important kernels which stress \gatherscatter performance.
} 

In particular, 
this work considers mini-apps from CORAL and CORAL-2, including AMG~\cite{osti_1389816}, LULESH~\cite{karlin2013lulesh}, and Nekbone~\cite{fischer2013nekbone}.
We built these mini-apps targeting ARMv8-A with support for Arm's Scalable Vector Extension (SVE)~\cite{7924233} at a vector length of 1024 bits.
The resulting executables were run through an instrumented version of the QEMU functional simulator~\cite{bellard2005qemu} to extract traces of all instructions accessing memory along with their associated virtual addresses.
From this instruction stream, collected from rank 0, we examine only \GS instructions, and extract the base address and offset vector for each, along with their frequencies.
The problem sizes are chosen so as to prioritize a realistic working set with 64 MPI ranks per node with one thread per rank, while the number of iterations is less emphasized.
For these apps, it is expected that multiple kernel iterations will have many patterns in common. More information on how we configured these codes is found in \cref{table:miniapp-details}.

\Cref{table:miniapp-char} shows the \gatherscatter characteristics extracted from several kernels selected from the aforementioned mini-apps, along with the amount of data motion performed by these \gatherscatter operations. 
The reported \GS data motion percentages are conservative, as current data records all scalar loads and stores them as being 64 bits, while a significant fraction of 32-bit scalar data types is expected.

Examination of the \gatherscatter behavior results in the observation of a small number of common \GS pattern classes:
\emph{uniform-stride}, where each element of a gather is a fixed distance from the preceding element;
\emph{broadcast}, where some elements of a gather share the same index;
\emph{mostly stride-1}, in which some elements of a gather are a single element away from the preceding element; and \emph{more complex strides}, in which elements of a gather have a complicated pattern containing many different strides.

We can make a few high-level remarks about \cref{table:miniapp-char}.
First, gathers are more common than scatters.
Secondly, \gatherscatter can account for high fractions of total load / store traffic (last column; up to 67.6\%, or just over two-thirds, in these examples).
Thirdly, the appearance of differing categories of stride types suggests that there are multiple opportunities for runtime (inspector / executor) and hardware memory systems to optimize for a variety of \gatherscatter use-cases, which Spatter can then help evaluate. 





\newcolumntype{b}{X}
\newcolumntype{s}{>{\hsize=.5\hsize}X}

\begin{table*}[ht]
\centering
\caption{Details for Selected Applications and Kernels Used for \GS Pattern Extraction.}
\label{table:miniapp-details}
\begin{small}
\begin{tabularx}{\textwidth}{sbb}
\toprule
   \textbf{Application} -- Version & Problem Size / Changes & Kernel Notes \\
\midrule
\textbf{AMG} -- \texttt{github.com/ LLNL/AMG} commit \texttt{09fe8a7} & Arguments \texttt{-problem 1 -n 36 36 36 -P 4 4 4}, also \texttt{mg\_max\_iter} in \texttt{amg.c} set to 5 to limit iterations. & Entirety of each of the functions listed in \Cref{table:miniapp-char}. \\
\textbf{LULESH} -- 2.0.3 & Arguments \texttt{-i 2 -s 40}, also modifications to vectorize the outer loop of the first loop-nest in \texttt{IntegrateStressForElems}. & The first loop-nest in \texttt{IntegrateStressForElems}. Arrays \texttt{[xyz]\_local[8]} as well as \texttt{B[3][8]} give stride-8 and stride-24. Also, the entirety of the  \texttt{InitStressTermsForElems} function. \\
\textbf{Nekbone} -- 2.3.5 & Set \texttt{ldim = 3}, \texttt{ifbrick = true}, \texttt{iel0 = 32}, \texttt{ielN = 32}, \texttt{nx0 = 16}, \texttt{nxN = 16}, \texttt{stride = 1}, internal \texttt{np} and \texttt{nelt} distribution. Also, \texttt{niter} in \texttt{driver.f} set to 30 to limit CG iterations. & First loop in \texttt{ax} (essentially a wrapped call to \texttt{ax\_e}) contains the observed stride-6. \\
\textbf{PENNANT} -- 0.9   & Config file \texttt{sedovflat.pnt} with \texttt{meshparams 1920 2160 1.0 1.125} and \texttt{cstop 5}. & Entirety of each of the functions listed in \Cref{table:miniapp-char}. \\
\bottomrule
\end{tabularx}
\end{small}
\end{table*}

\subsection{Open Source Techniques for Pattern Analysis}
The application analysis done in this work depends on a custom, closed-source QEMU functional simulator that uses an SVE vector size of 1024 and data from the first rank of an MPI run, which may slightly differ from other ranks. along with post-processing scripts to extract the most utilized \gatherscatter patterns. We recognize this is a limitation of the current work in that trace capture and analysis can be one of the most time-consuming portions of an analysis of \gatherscatter patterns.  

A useful open-ended project that would improve pattern inputs for not just Spatter but many other application analysis frameworks would be a tool based on either DynamoRio (which supports AVX-512 and SVE instructions) or Intel Architecture Code Analyzer (IACA) that generates this type of data in a less time-consuming fashion and that performs correlation and clustering across all ranks of an application for CPU and GPU codes.

%% file: sections/benchmark_design.tex
\section{Design of the Spatter Benchmark} \label{sec:bmk_design}

\begin{figure}
\centering
\includegraphics[width=0.95\linewidth]{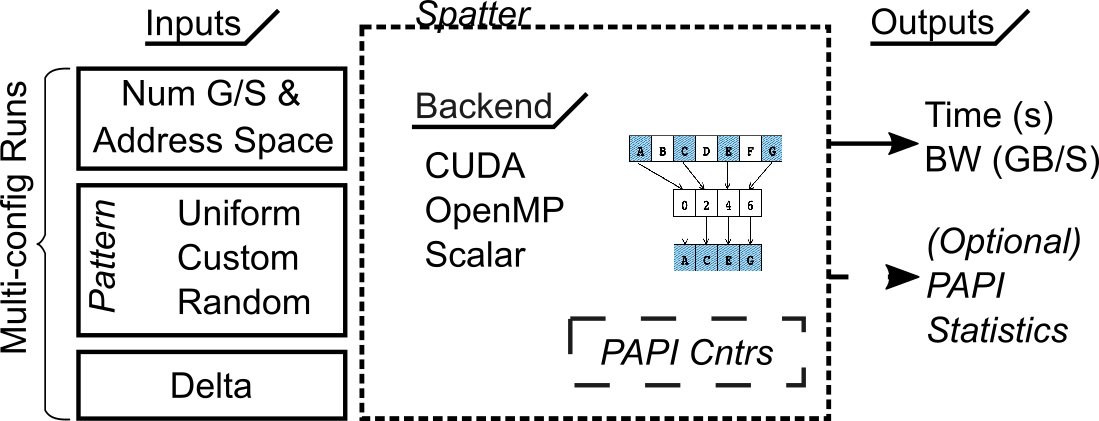}
	\caption{An overview of the Spatter benchmark with inputs and outputs.}
\label{fig:spatter-overview}
\end{figure}

We have developed Spatter because existing benchmark suites like STREAM\cite{mcalpin:2018:streamingstore} and STRIDE\cite{seager:2019:stridebmk} focus on uniform stride accesses and are not configurable enough to handle non-uniform, indirect accesses or irregular patterns. For more information on related benchmarks, see \cref{sec:related-work}. \Cref{fig:spatter-overview} shows a conceptual view of the Spatter benchmark. The design of the benchmark is described further below. 

\begin{algorithm}
    \caption{Gather Kernel}
    \label{alg:gather}
    \begin{algorithmic}
        \FOR{i in 1..N}
            \STATE{\texttt{src = src + delta * i}}
            \FOR{j in 1..vector\_length}
                \STATE{\texttt{dst[j] = src[idx[j]]}}
            \ENDFOR
        \ENDFOR
    \end{algorithmic}
    \algcomment{The basic gather algorithm. Scatter is performed analogously. False sharing is prevented by giving each thread its own \texttt{dst} buffer for gather, and \texttt{src} buffer for scatter. }
\end{algorithm}

\subsection{Kernel Algorithm}
Spatter represents a memory access pattern as a short index buffer, and a delta. At each base address address \texttt{delta*i}, a gather or scatter will be performed with the indices in the index buffer.
The pseudocode is in \Cref{alg:gather}, and a visual representation is in \Cref{fig:spatter-algo}. This algorithm allows us to capture some spatial and temporal locality: spatial locality can be controlled by choosing indices that are close together, and temporal locality can be controlled by picking a delta that causes your gathers to overlap. In either case, the locality will be fixed for the entirety of the pattern. Implementation details for the different backends are described in the following section.

\subsection{Backends} 
\begin{figure}
\centering
\includegraphics[width=0.95\linewidth]{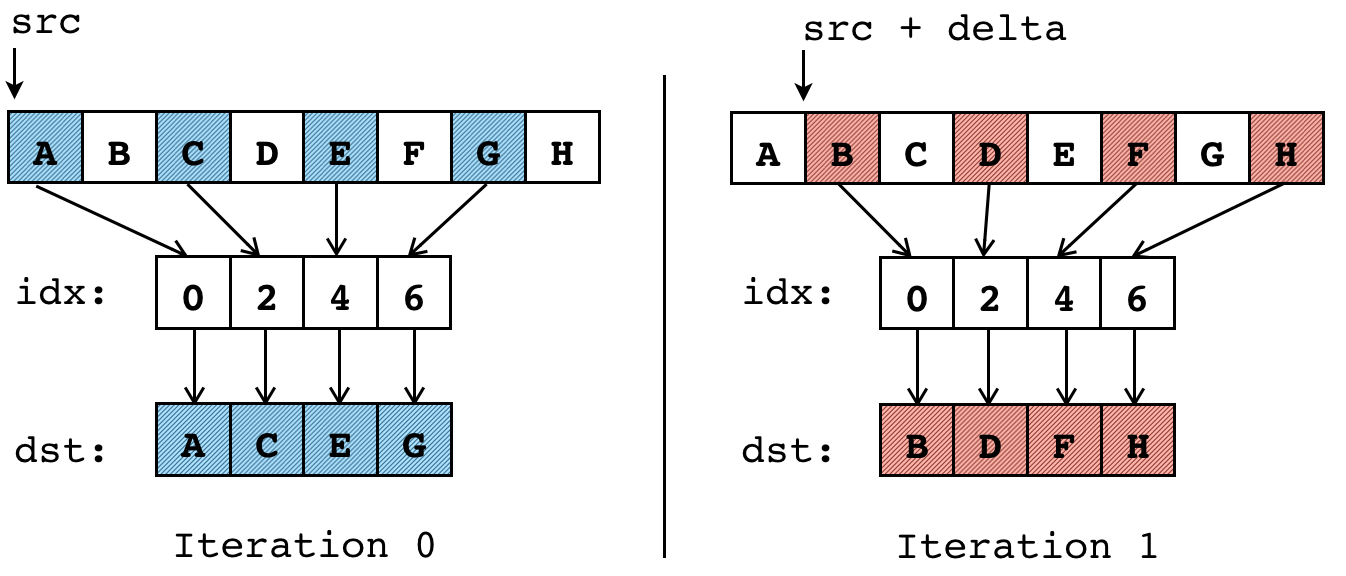}
\caption{\small A visual representation of the first two iterations of the gather kernel with a uniform stride-2 index buffer of length 4, and a delta of 1.}
\label{fig:spatter-algo}
\end{figure}

Spatter contains Gather and Scatter kernels for three backends: Scalar, OpenMP, and CUDA. A high-level view of the gather kernel is in \Cref{fig:spatter-algo}, but the different programming models require that the implementation details differ significantly between backends. Spatter provides performance tuning knobs for both the OpenMP and CUDA backbones, such as index buffer length and block size. 

\textbf{OpenMP:} The OpenMP backend is designed to make it easy for compilers to generate \GS instructions. Each thread will perform some portion of the iterations shown in \Cref{fig:spatter-algo}. 
To ensure high performance when gathering, each thread will gather into a local destination buffer (vice-versa for scattering). This avoids the effects of false sharing. 

\textbf{CUDA:} Whereas in the OpenMP backend, each thread will be assigned its own set of iterations to perform, in the CUDA programming model, an entire thread block must work together to perform an iteration of \Cref{fig:spatter-algo} to ensure high performance. 
These backends are similar, in that each thread block gathers into thread local memory to allow for high performance. The major difference is that each thread block must read the index buffer into shared memory to achieve high performance on Spatter's indexed accesses. 

\textbf{Scalar:} The Scalar backend is based on the OpenMP backend, and is intended to be used as a baseline to study the benefits of using CPU vector instructions as opposed to scalar loads and stores.
The major difference between this and the OpenMP backends is that the Scalar backend includes a compiler pragma to prevent vectorization, namely \texttt{\#pragma novec}.

\subsection{Benchmark Input}
The power of Spatter comes from its ability to easily express many different memory patterns of interest to us. Spatter accepts either a single index buffer and run configuration as input, or a JSON file containing many such patterns and configurations. 

\textbf{Pattern Specification:}  Spatter currently provides robust mechanisms for representing spatial locality with both standard  patterns and more complex, custom patterns for representing indirect accesses. In Spatter, a memory access pattern is described by specifying (1) either gather or scatter (2), providing a short index buffer, (3) a delta, and (4) the number of gathers or scatters to perform. Spatter will determine the amount of memory required from these inputs. Spatter includes three built-in, parameterized patterns, which are Uniform Stride, Mostly Stride-1, and Laplacian.  These all describe small index sets, which should be thought of as the offsets for a single \GS. When combined with a delta, these will describe a memory access pattern. They are described in further detail below. 

\subsubsection{Uniform Stride}
The Uniform Stride index buffer is specified to Spatter with \texttt{UNIFORM:N:STRIDE}. It generates an index buffer of size \texttt{N} with stride \texttt{STRIDE}. For example, the index buffer generated by \texttt{UNIFORM:8:4} is \texttt{[0,4,8,12]}. 
\subsubsection{Mostly Stride-1}
The Mostly Stride-1 index is the result of accessing a few elements sequentially, and then making some jump and accessing a few more elements sequentially. In code, this could be the result of accessing the same few members of structs in an array, or from accessing a sub-block of a matrix. In Spatter, you can specify an MS1 pattern with \texttt{MS1:N:BREAKS:GAPS}. The pattern will be length \texttt{N}, with gaps at \texttt{BREAKS}, with gaps of size \texttt{GAPS}. For example, the index buffer generated by \texttt{MS1:8:4:20} is \texttt{[0,1,2,3,23,24,25,26]}.
\subsubsection{Laplacian}
The Laplacian index is based on Laplacian stencils from PDE solvers.  Spatter can generate 1-D, 2-D, or 3-D stencils with the pattern \texttt{LAPLACIAN:D:L:SIZE}. This creates a \texttt{D}-dimensional stencil, with each ``branch'' of length \texttt{L}, for a problem size of \texttt{SIZE}. Even though a 2- or 3-D problem can be specified, Spatter still only allocates a 1-D data array. Thus the problem size must be specified in the stencil so that Spatter can calculate the distances of the neighbors in the stencil. For example, the input \texttt{LAPLACIAN:2:2:100} generates the classic 5-point stenci \texttt{[0,100,198,199,200,201,202,300,400]}, which may be more familiar to users in it's non-zero-based form, \texttt{[-200,-100,-2,-2,0,1,2,100,200]}. 


\subsubsection{Custom Patterns (Complex Accesses)}
Finally, if users want to represent a more complex pattern not specified above, they can specify a pattern index buffer as \texttt{./spatter -p idx0,idx1,...,idxN}. This allows users to develop more complex and irregular kernels that often show up in HPC applications. The use of custom patterns is the basis of ongoing research described in \cref{sec:futurework}.

\textbf{Delta Specification:}
To form a full memory access pattern, Spatter needs an index buffer, as described in , and a delta. The index buffer will be used as the offsets for a gather or scatter with base addresses \texttt{0, delta, 2*delta} etc.

\textbf{JSON Specification:} When running tests, it is common to run many different patterns. To support this, Spatter accepts a JSON file as input that can contain as many configurations as the user wishes. Spatter will parse this file and allocate memory once for all tests, greatly speeding up test sets with many different patterns, and easing data management.

\subsection{Example}
With an index buffer, a delta, and the numbers of gathers or scatters, we are ready to specify a Spatter run. The simplest example would be to emulate a STREAM-like pattern, which would look like a uniform stride-1 pattern with delta equal to the index length, so that there is no data reuse between gathers. In Spatter, this is expressed as:

\footnotesize
\begin{verbatim}
./spatter -k Gather -p UNIFORM:8:1 -d 8 -l $((2**24)) 
\end{verbatim}
\normalsize

to run $2^{24}$ (\texttt{-l}) gathers (\texttt{-k}), each one 8 doubles beyond the last (\texttt{-d}), and each using an index buffer of length 8 and uniform stride-1 (\texttt{-p}).  This will produce a STREAM Copy-like number, but it will only be read bandwidth, as a gather reads data from memory to a register. Spatter includes further options for choosing backends and devices and performance tuning that are described in its \texttt{README}. 

\subsection{Benchmark Output}

For each pattern specified, Spatter will report the minimum time taken over 10 runs to perform the given number of gathers or scatters. It will also translate this into a bandwidth,  with the formula \texttt{Bandwidth = (sizeof(double) * len(index) * n) / time}, where \texttt{n} is the number of gathers or scatters. This is the amount of data that is moved to or from memory, and does not count the bandwidth used by the the index buffer, as it is assumed to be small and resident in cache. We concede that this number may not be a true bandwidth in the traditional sense of the word, as many patterns specified to Spatter may allow for cache reuse. Thus, one should think about the bandwidths reported as the rate at which the processor is able to consume data for each pattern.

Optionally, PAPI \cite{terpstrapapi} can be used to measure performance counters, but we do not demonstrate that feature in this paper. 

For JSON inputs, Spatter will also report stats about all of the runs, such as the maximum and minimum bandwidths observed across configurations, as well as the harmonic mean of the bandwidths.

%% file: sections/exp_setup.tex
\section{Experimental Setup} \label{sec:exp}

\begin{table*}[ht]
\centering
\caption{Experimental Parameters and Systems (OMP Denotes OpenMP, and OCL Denotes OpenCL).}
\label{table:exp-systems}
\begin{tabular}{lccrc}
\toprule
   System description &   Abbreviation & System Type & STREAM (MB/s)  & Threads, Backends\\
\midrule
  Knight's Landing (cache mode) & KNL &  Intel Xeon Phi & 249,313 & 272 threads, OMP\\
  Broadwell & BDW &   32-core Intel CPU (E5-2695 v4) & 43,885 & 16 threads, OMP\\
  Skylake & SKX &  32-core Intel CPU (Platinum 8160) & 97,163  & 16 threads, OMP\\
  Cascade Lake & CLX &  24-core Intel CPU (Platinum 8260L) & 66,661  & 12 threads, OMP\\
  ThunderX2 & TX2 &  28-core ARM CPU & 120,000 & 112 threads, OMP\\
  Kepler K40c & K40c &  NVIDIA GPU & 193,855 & CUDA\\
  Titan XP & Titan XP &  NVIDIA GPU & 443,533 & CUDA\\
  Pascal P100 & P100 &  NVIDIA GPU & 541,835 & CUDA\\
  Volta V100 & V100 &  NVIDIA GPU & 868,000 & CUDA\\
\bottomrule
\end{tabular}
\end{table*}

\cref{table:exp-systems} describes the  different configurations and backends tested for our initial evaluation using the Spatter benchmark suite.
We pick a diverse set of systems based on what is currently available in our lab and collaborator's research labs, including a Knight's Landing system, and a prototype system with ARMv8 ThunderX2 chips designed by Cavium.
We also include a server-grade and desktop-grade Intel CPU system and several generations of NVIDIA GPUs. Unfortunately we currently do not have access to a recent AMD GPU, CPU, or APU system for testing but hope to include this in future experiments. 

\textbf{OpenMP: } To control for NUMA effects, CPU systems are tested using all the cores on one socket or one NUMA region if the system has more than one CPU socket. Some systems like the KNL on Cori have an unusual configuration where the entire chip is listed as 1 NUMA region with 272 threads. For all the OpenMP tests, Spatter is bound to one socket and run using one thread per core on that socket. The following settings are used for OpenMP tests: 
\begin{inparaenum}
\small
\item \verb-OMP_NUM_THREADS = <num_threads_single_socket>-
\item \verb-OMP_PROC_BIND = master- \\
\item \verb-OMP_PLACES = sockets- \\
\item \verb-KMP_AFFINITY = compact- (only for KNL)
\end{inparaenum}

An important performance tuning factor is the index buffer length. On CPUs, we found that it was best to use an index buffer that closely matches the hardware vector length, or a small multiple. On the CPUs we tested, we achieved good performance by using an index buffer length of 16, which is 2-4x the length of the vector registers on our systems. 

\textbf{CUDA: }
When testing on GPUs, the block size for Spatter is set at 1024 and an index buffer of length 256 is used. These settings allow Spatter to reach bandwidths within 20\% of the vendor reported theoretical peak for both gather and scatter kernels.
These bandwidths are slightly different than what is typically reported, as gather is designed to only perform reads, and scatter should only perform writes.

\textbf{Experimental Configurations: } Runs of Spatter use the maximum bandwidth out of 10 runs for the platform comparison uniform stride and application pattern tests.
STREAM results are generated using $2^{25}$ elements with either STREAM for CPU or BabelStream for GPU, while all Spatter uniform stride tests read or write at least 8GB of data on the GPU and 16GB on the CPU. The difference between CPU and GPU data sizes results from most GPUs having less than 16 GB of on-board memory. The application-specific pattern tests read or write at least 2GB. 



%% file: sections/results.tex
\section{Case Studies} \label{sec:results}
Spatter is designed to be a flexible tool that can allow the user to run many different memory access patterns and expose many knobs used for tuning. In this section, we use Spatter to investigate several questions regarding CPU and GPU memory architecture including: \textit{A)} uniform stride access on CPUs, \textit{B)} uniform stride access on GPUs, \textit{C)} the effectiveness of gather/scatter over scalar load/store, and \textit{D)} the performance of trace-derived gather/scatter patterns on CPU. 

\subsection{CPU Uniform Stride}

\begin{figure}[!htpb]
\centering
\begin{tabular}{c}
\includegraphics[width=0.68\linewidth]{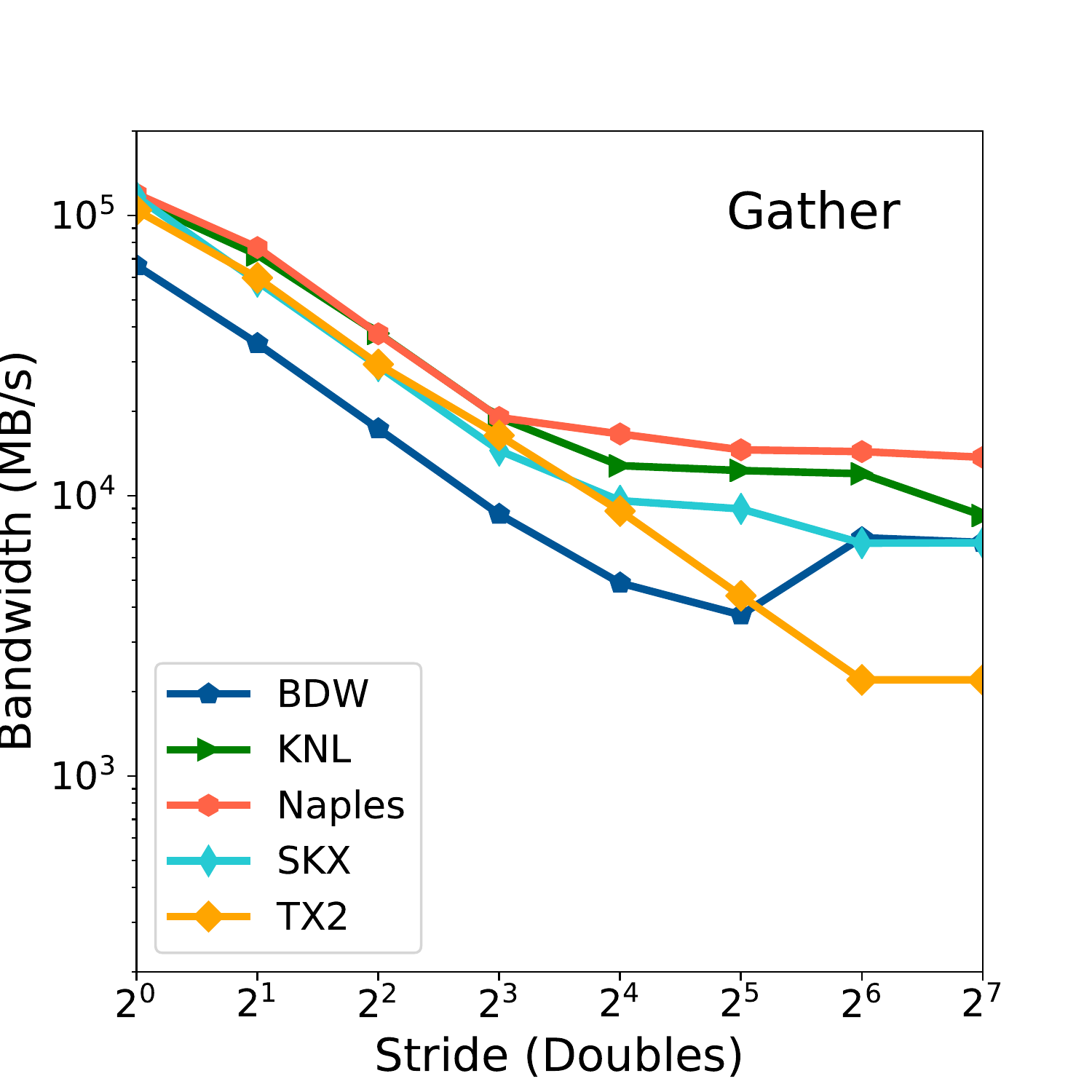} \\
\textbf{(a)} \\
\includegraphics[width=0.68\linewidth]{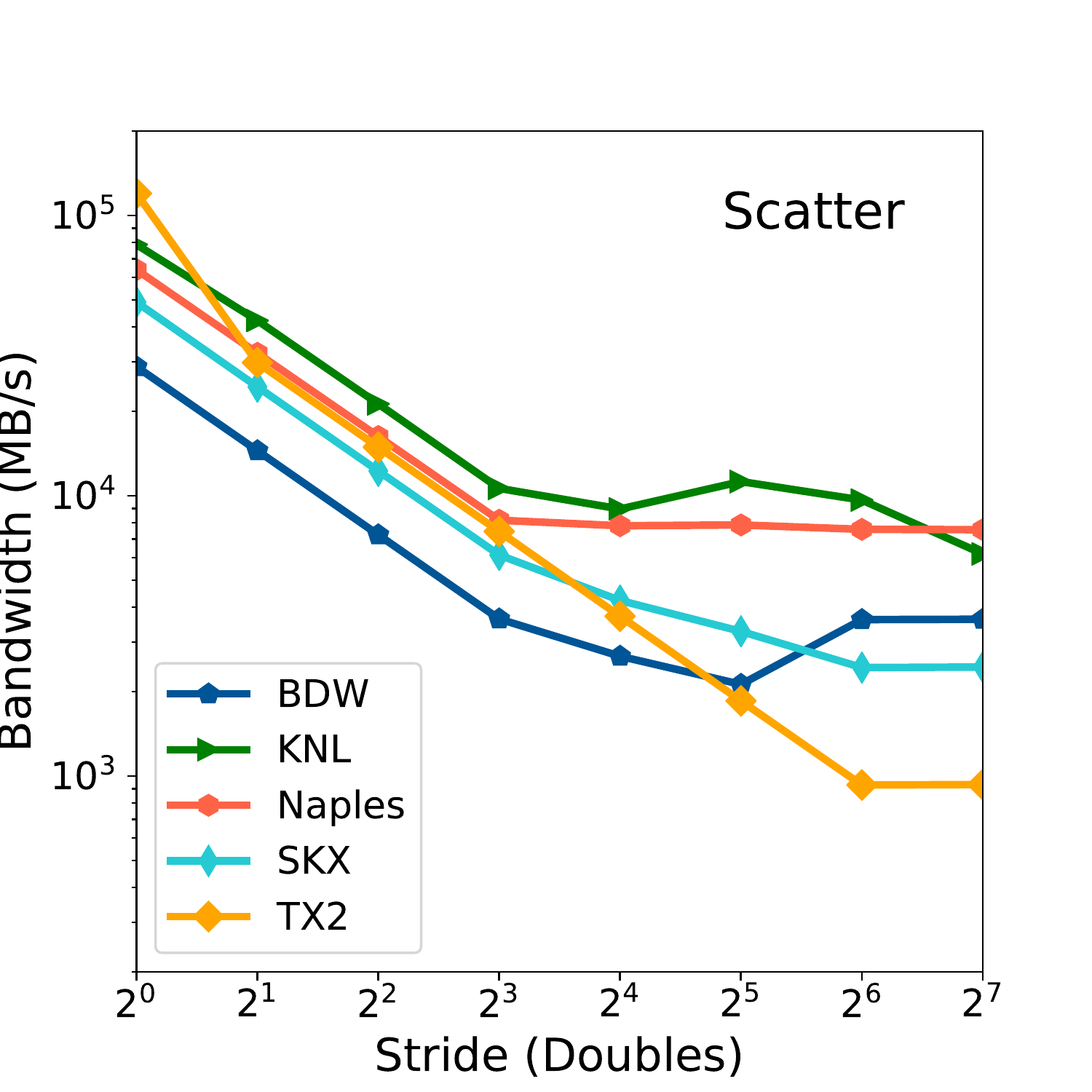} \\
\textbf{(b)} \\
\end{tabular}
\captionof{figure}{\small CPU Gather and Scatter Bandwidth Comparison \textmd{\\We increase the stride of memory access and show how performance drops as the stride increase from 1 to 128 doubles on Skylake, Broadwell, Naples, and Thunder X2 systems. Cascade Lake is omitted as it overlaps closely with Skylake. A log-scale is used for the y-axis to make differences between the platforms apparent.} \textit{Takeaway: Uniform stride patterns show us that peak bandwidth is not necessarily an indication of which architecture will perform best at even moderate strides.}}
\label{fig:cpu_ustride}
\label{fig:cpu_ustride_scatter}
\end{figure}

We start our investigation by performing a basic test: running Spatter with the uniform stride pattern, and increasing the stride by 2x until performance flattens. At stride 1, this is analogous to the STREAM benchmark\footnote{On CPU, we use an index buffer of length 8 and fill it with indices [1*stride, 2*stride, ...]. We set the delta to be 8*stride, so that there is no data reuse and indeed stride-1 matches the STREAM pattern.}, except for the fact that Spatter will only generate read instructions (gathers) for the gather kernel and write instructions for the scatter kernel, meaning the bandwidths should be slightly different. \cref{fig:cpu_ustride} shows the results of our uniform stride tests on CPUs. We would expect that as stride increases by a factor of 2, bandwidth should drop by half; the entire cache line is read in but only every other element is accessed. This should continue until about stride 8, as we are then using one double from every cache line. This is what we see on Naples, but performance continues to drop on TX2, Skylake, and Broadwell. Interestingly, Broadwell performance increases at stride-64, even out-performing Skylake. We can further use Spatter to investigate these two points: 1) Why does Broadwell outperform Skylake at high strides, and 2) why does TX2 performance drop so dramatically past 1/16?

\subsubsection{Disabling Prefetching}
\begin{figure}[!t]
\vskip .55cm
\centering
\begin{tabular}{c}

\includegraphics[width=0.75\linewidth]{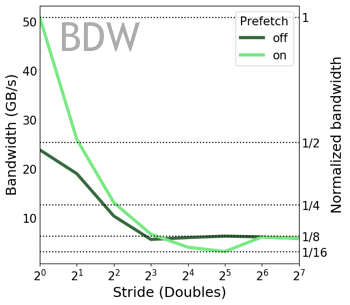}\\
\textbf{(a)}\\
\includegraphics[width=0.75\linewidth]{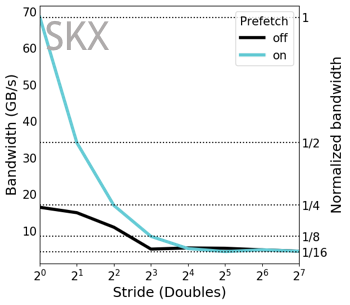}\\
\textbf{(b)}\\
\end{tabular}

\caption{{\small Broadwell Gather (a) and Skylake Gather (b) \textmd{ We show the performance of gather for various strides, with prefetching on and off. On the right, normalized bandwidth is shown to display the regularity of the decrease in bandwidth. }}\textit{Takeaway: Uniform stride patterns can help us identify interesting prefetching behavior, such as above, where we see that Skylake always fetches two lines.}}
\label{fig:cpu_ustride_ind}
\end{figure}

To get an idea of what is causing Broadwell to outperform Skylake, we turn prefetching off with Model Specific Registers (MSRs) and re-run the same uniform stride patterns. \cref{fig:cpu_ustride_ind}a and b shows the results from this test. For Broadwell, performance doesn't show the same increase for stride-64 with prefetching off and it instead bottoms out after stride-8. We conclude that one of Broadwell's prefetchers pulls in two cache lines at a time for small strides but switches to fetching only a single cache line at stride-64 (512 bytes). We can understand the performance discrepancy between Broadwell and Skylake by looking at \cref{fig:cpu_ustride_ind}b. Performance drops to 1/16th of peak, as Skylake always brings in two cache lines, no matter the stride. We did not get the opportunity to run on the Thunder X2 without prefetching since it does not have a similar MSR equivalent, but we suspect similar effects are at play: one of the prefetchers likely always brings in the next line, although that only helps to explain performance dropping through stride-16, not through stride-64. 

\textit{Lesson: By running uniform stride tests on CPUs we are able to (1) identify a number of performance crossover points for intermediate strides, and (2) see some interesting differences between the Broadwell and Skylake prefetching behavior}

\subsection{GPU Uniform Stride}

\begin{figure}[!t]
\centering
\begin{tabular}{c}

\includegraphics[width=0.68\linewidth]{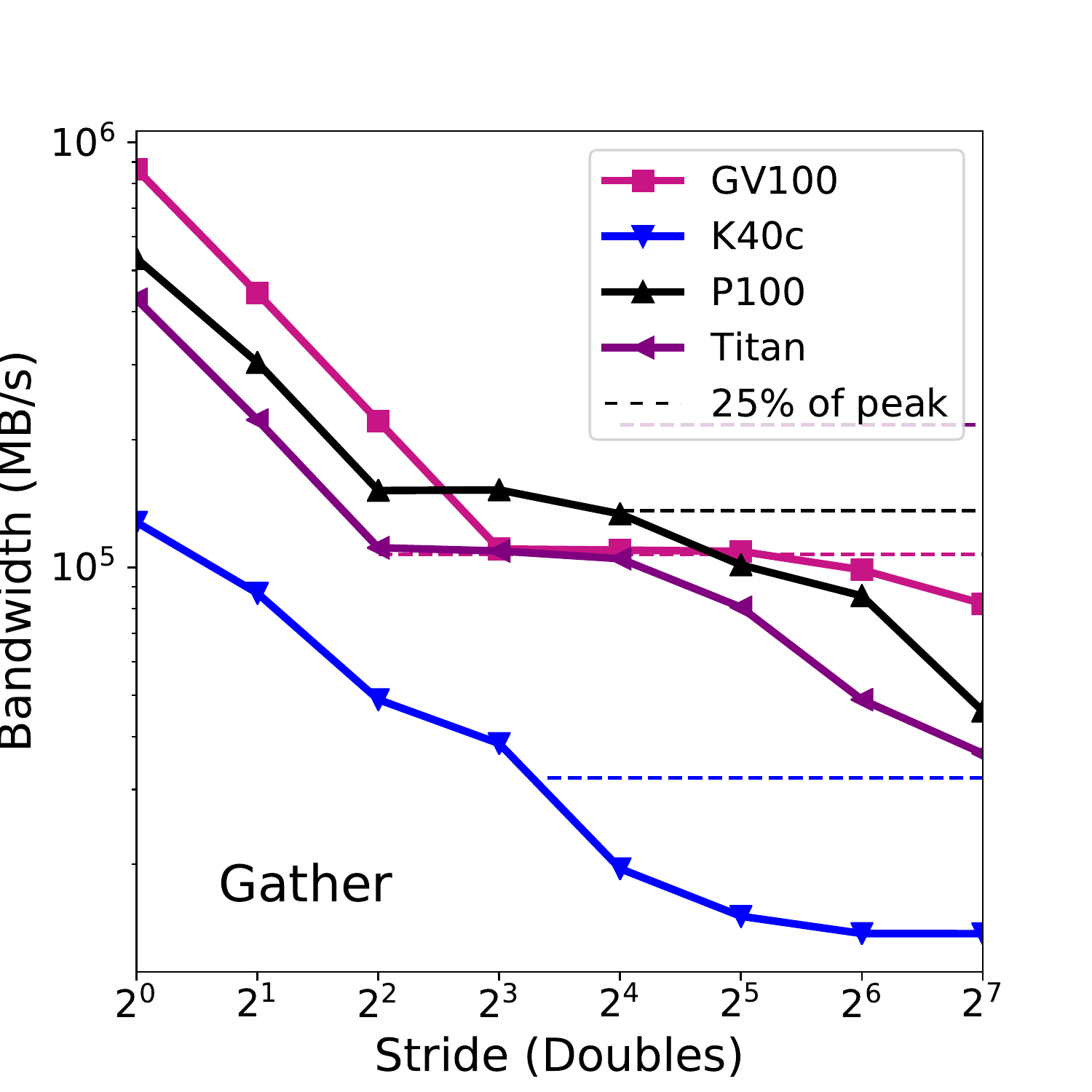}\\
\textbf{(a)}\\
\includegraphics[width=0.68\linewidth]{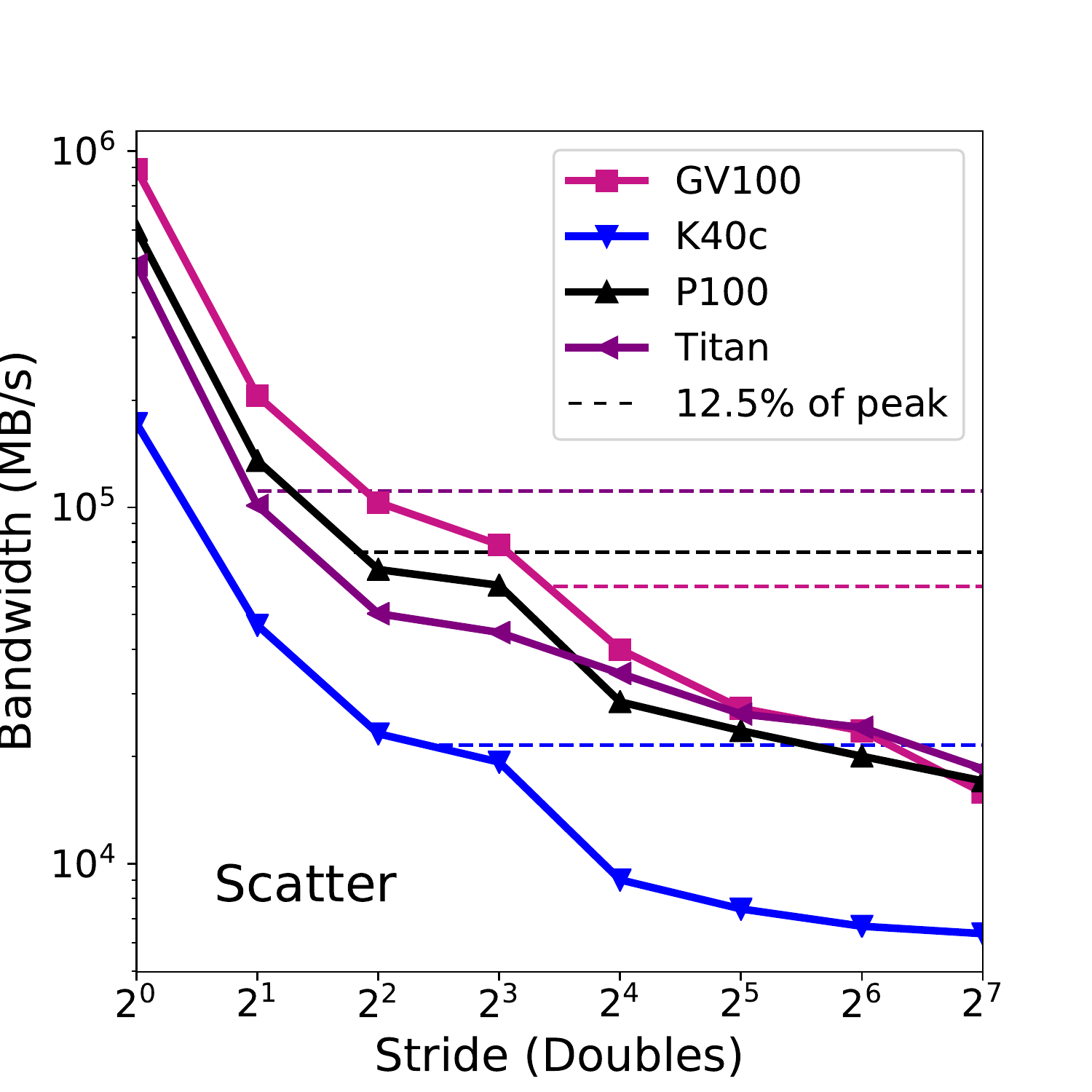} \\ 
\textbf{(b)}\\
\end{tabular}

\caption{\small GPU Gather (a) and Scatter (b) Uniform Stride Bandwidth comparison\textit{ Takeaway: We are able to use uniform stride patterns to show improvements to the memory architecture of GPUs over time, beyond simply improved bandwidth. We see here that in newer generations, not only do GPUs have more bandwidth, they are also able to utilize a higher percentage of that bandwidth throughout intermediate strides.}}
\label{fig:gpu_ustride}
\end{figure}

As the memory architecture of CPUs and GPUs is quite different, it is worthwhile to see how GPUs handle these uniform stride patterns. \cref{fig:gpu_ustride}a shows how a K40c, a Titan Xp, and a P100 perform on the same tests\footnote{To get high performance on GPUs, the threads within a block all work together to read a pattern buffer into shared memory This buffer must be much longer than the CPU index buffer (256 indices vs 8) so that each thread has enough work to do.}. As with the CPUs, we see bandwidth drop by half for stride-2 and by another half for stride-4.However, for the P100 and the Titan Xp, from stride-4 to stride-8, we see that bandwidth stays the same (illustrated by the dotted lines). This is due to the fact that GPUs are able to coalesce some loads. The older K40 hardware shows less ability to do so. In the scatter kernel plot, \cref{fig:gpu_ustride}b, the effect of coalescing is less pronounced, but still visible from stride 4 to stride 8. Instead of plateauing at 1/4th of peak bandwidth, however, it plateaus at 1/8th. Regardless of the effect being less pronounced in scatter vs. gather, we still see the benefit of a memory architecture that is able to coalesce access, as we see how the bandwidth curves of these GPUs platforms take a longer time to fall off than their CPU counterparts. 

\textit{Lesson: By running uniform stride tests on GPUs, we identify some qualitative differences between CPU and GPU strided access, especially in the range of stride-8 to stride-32.}
 

 
\begin{figure}[!t]
\centering
\begin{tabular}{c}

\includegraphics[width=0.68\linewidth]{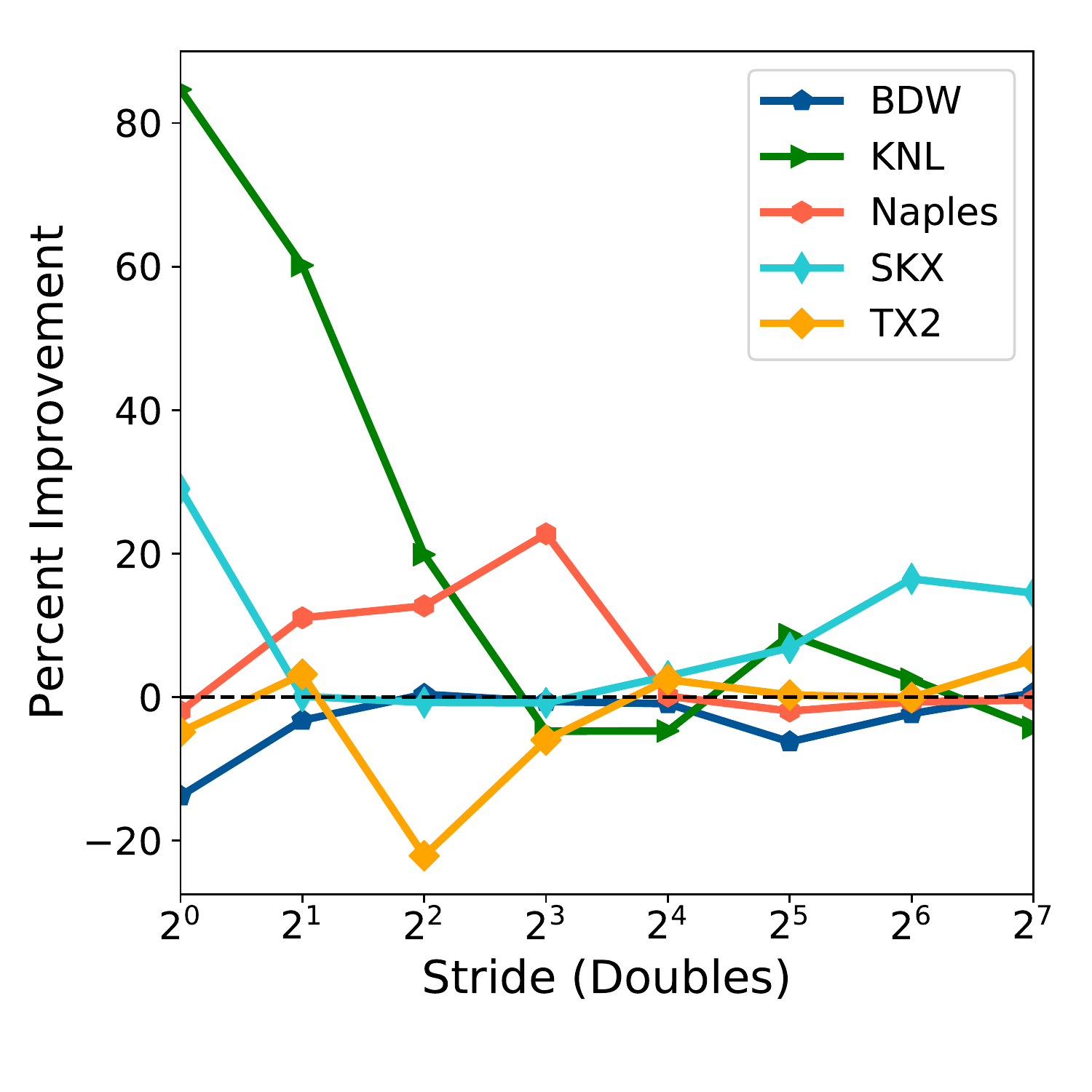}\\
\textbf{(a)}\\
\includegraphics[width=0.68\linewidth]{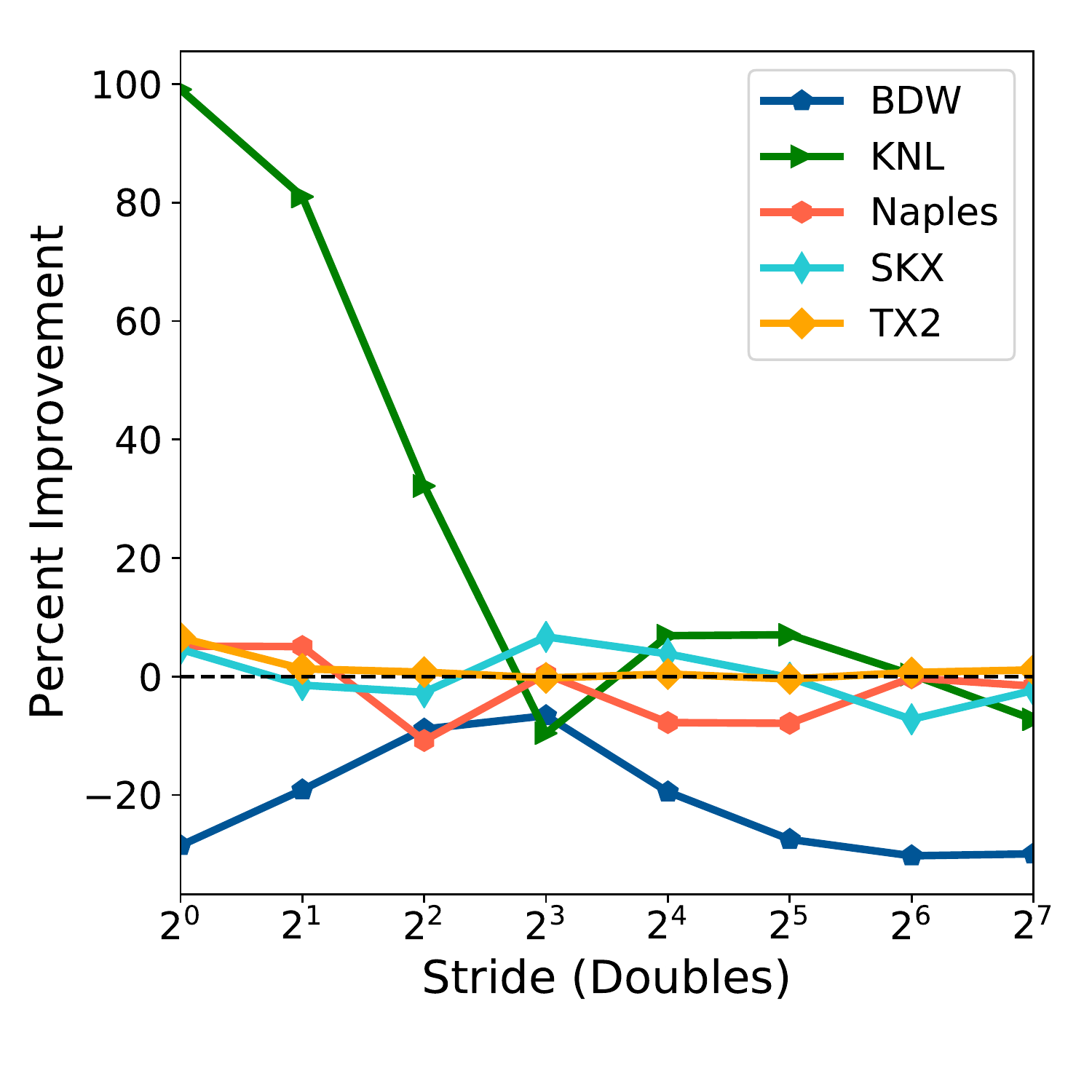}\\
\textbf{(b)}\\
\end{tabular}

\caption{\small Improvement of SIMD Gather Kernel (a) and Scatter Kernel (b) Compared to Serial Scalar Backend.\textit{Takeaway: By examining the performance of uniform stride patterns with and without vectorization enabled in the compiler, we show that there achieving maximum bandwidth processors such as Knights Landing and Skylake requires vectorization. On the other hand these instructions can be detrimental to performance on Broadwell.}}
\label{fig:spatter-vector-speedup}
\end{figure}

\subsection{SIMD vs. Scalar Backend Characterization}
\label{ssec:scalar_simd}

Spatter can also be used to test the effectiveness of different hardware implementations of single instruction, multiple data (SIMD) instruction set architectures (ISAs). In a real-world sense, this capability can be used by compiler writers to answer questions such as whether it would be beneficial to load some addresses with vector instructions and others with scalar instructions to maximize utilized memory bandwidth. Vector versions of indexed load and store instructions help compilers to vectorize loops and can also help avoid unnecessary data motion between scalar and vector registers that might otherwise be required. We can use Spatter to investigate whether these vector instructions are indeed superior to scalar load instructions and whether compiler writers should prioritize vectorized \gatherscatter optimizations.

To demonstrate the effectiveness of SIMD load / store instructions, we run Spatter using the gather kernel on multiple platforms with the scalar backend as a baseline.
This scalar baseline is then compared to the OpenMP backend as vectorized by the Cray compiler (CCE 9.0) with the resulting percent improvement from vectorization reported in Figure \ref{fig:spatter-vector-speedup} for strides 1-128, as before.
The Broadwell CPU performs the worst of all the tested CPUs, showing worse performance with vectorized code in many cases for both Gather and Scatter. Thus, for a memory heavy kernel, it would likely be better to use scalar instructions than \GS instructions, although this difference may be mitigated as \GS instructions remove the need to move data between regular and vector registers.

On the other hand, Skylake, Knights Landing, and Naples have better gather performance in the vectorized case.  The use of gather instructions on these platforms is clearly justified. Of these three, however, Naples is the only one to not improve in the scatter case as well. This is due to the lack of scatter instructions on Naples. TX2 has no \GS support at all, so it stays close to 0\% difference (save for a single outlier in the gather chart). Interestingly, for our three processors with useful \GS instructions, they all gather best in different regions, with Knights Landing best at small strides, Naples for medium strides, and Skylake best at large strides. While we are not able to explain the reason for this performance artifact, we have demonstrated the benefit of \GS instructions over their scalar counterparts. At least for Knights Landingn, anecdotal evidence has suggested that using vectorized instructions at lower strides reduces overall unique instruction count and overall request pressure on the memory system.

\textit{Lesson: Spatter shows that the \GS instructions themselves lead to higher performance, especially on bottlenecked platforms like Knight's Landing. \GS instructions have a further benefit over scalar in that the data loaded is already in a vector register, whereas after performing scalar loads, further rearrangement would be needed to move the data into vector registers. }

\subsection{Application-derived \GS Patterns}

While the three previous sections have focused on uniform stride patterns, Spatter is also able to run more complex patterns. To demonstrate Spatter's ability to emulate patterns found in real applications, we take the top patterns from several DoE mini-apps (as described in \cref{sec:app_char}) and run them in Spatter. The patterns that come out of \cref{sec:app_char} are described by a buffer of offsets and a delta. These offsets and deltas can be found in \cref{table:patterns} which is in \cref{sec:pattern_appendix}.

In \cref{sec:covariance}, we first look at how these patterns perform in aggregate, and see if they correlate with STREAM bandwidth. In \cref{sec:radar}, we look at each pattern individually, and look for trends among the applications. Finally, in \cref{sec:bwbw}, we show a method for plotting results that allows us to examine absolute and relative performance of patterns at the same time. 

\subsubsection{Application Pattern / STREAM Correlation}
\label{sec:covariance}

To examine the patterns, we first take an aggregate look at the performance differences between CPUs and GPUs. \cref{table:app-trace} shows the harmonic mean of the performance of the patterns. To see if the performance correlates with STREAM, we calculate Pearson's correlation coefficient, R, as follows: 
\begin{equation}
\mathrm{R}=\mathrm{cov}\!\left(X, STREAM\right)/\left(\mathrm{std}\!\left(X\right)*\mathrm{std}\!\left(STREAM\right)\right)
\end{equation}

What we note from \cref{table:app-trace} is the following: in aggregate, LULESH shows poor performance on most CPU platforms. We will see in the next section that this is due to pattern LULESH-S3, which is a scatter with delta 0. We believe this triggers cache invalidations for multi-core writebacks. Next, we see that AMG and Nekbone show higher performance than STREAM in general. This is due to the effects of caching for the included patterns.

More interestingly, we see that the CPU runs of the Nekbone and PENNANT patterns show poor correlation (close to 0) with STREAM. In the case of AMG, the patterns perform much better than STREAM, whereas in PENNANT, the patterns perform much worse. This means that Spatter is indeed capturing different behavior than STREAM, and that the patterns Spatter generates are not well approximated by STREAM on CPUs. For GPU systems, however, the R coefficient shows that STREAM is well correlated (close to 1) with the Spatter results. As GPU caches are much less complex than CPUs, this is to be expected. 


\begin{table*}{}
\centering
\caption{Spatter Results for Mini-apps}
\label{table:app-trace}
\begin{small}
\begin{tabular}{l|ccccc}
\toprule
   \textbf{Platform} & AMG \begin{scriptsize}(n=36)\end{scriptsize} & Nekbone \begin{scriptsize}(n=6)\end{scriptsize} & Lulesh & PENNANT & STREAM\\
  & GB/s & GB/s & GB/s & GB/s & GB/s\\
   & (H-Mean) &  (H-Mean) & (H-Mean) & (H-Mean) & \\
\midrule
\midrule
   \textbf{BDW}     & 123 & 121 & 20& 6  & 43\\
   \textbf{SKX}     & 328 & 309 & 12  & 35 & 96\\
   \textbf{CLX}     & 315 & 287 & 14   & 41 & 94\\
   \textbf{Naples}  & 140 & 323 &3 & 11 & 97 \\ 
   \textbf{TX2}     & 270 & 247 & 232 & 28& 241\\
   \textbf{KNL}     & 201  & 190  & 19  & 4 & 249\\
   \textbf{R-value}     & 0.15 & -0.04 & 0.50 & -.1&\\
\midrule
   \textbf{K40c}    & 108  & 99   & 88  & 14& 193 \\
   \textbf{TitanXP} & 496  & 320  & 175 &21 & 443 \\
   \textbf{P100}    & 703  & 673  & 165 & 19& 541 \\
   \textbf{R-value}     & 0.66 & 0.62 & 0.62 &0.57 & \\

\bottomrule
\end{tabular}
\end{small}
\end{table*}

\subsubsection{Comprehensive Evaluation Across Platforms and Applications}
\label{sec:radar}

The design of Spatter makes it easy to collect lots of data, over many platforms and patterns, and these results are potentially more revealing than single-number benchmarks like STREAM as discussed in \Cref{sec:covariance}. A natural question is whether that data facilitates any qualitative comparisons about broad classes of platforms or applications. For example, what can we say generally about CPU-oriented memory system design today versus GPU-oriented design? Are applications uniform in their patterns, or are they more varied?

\begin{figure*}
  \includegraphics[width=.98\textwidth,height=.98\textheight,keepaspectratio]{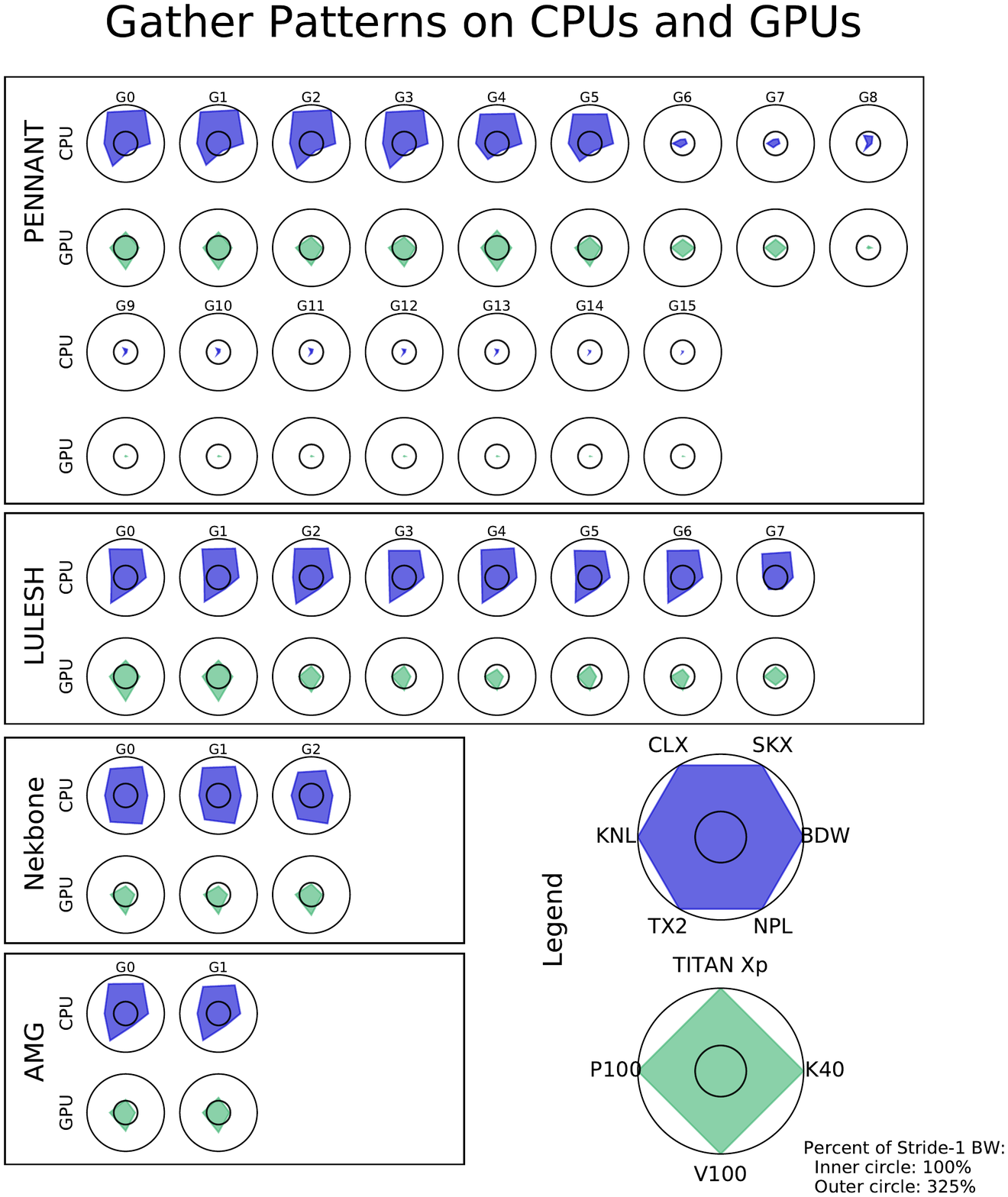}
  \vspace{-1cm}
  \caption{App-derived Gather Patterns}{Each circle represents the a single pattern. A spoke represents the performance of that pattern on a specific architecture as a percentage of the architecture's stride-1 bandwidth. The pattern descriptions are in \ref{table:patterns}.}{\textit{ Takeaway: As the chart report relative and not absolute performance, we must emphasize that this does not show CPUs outperforming GPUs, but rather the ability of CPUs to utilize their caches on the chosen patterns. }}
  \label{fig:radar_gather}

\end{figure*}
\begin{figure}
\includegraphics[width=0.5\textwidth]{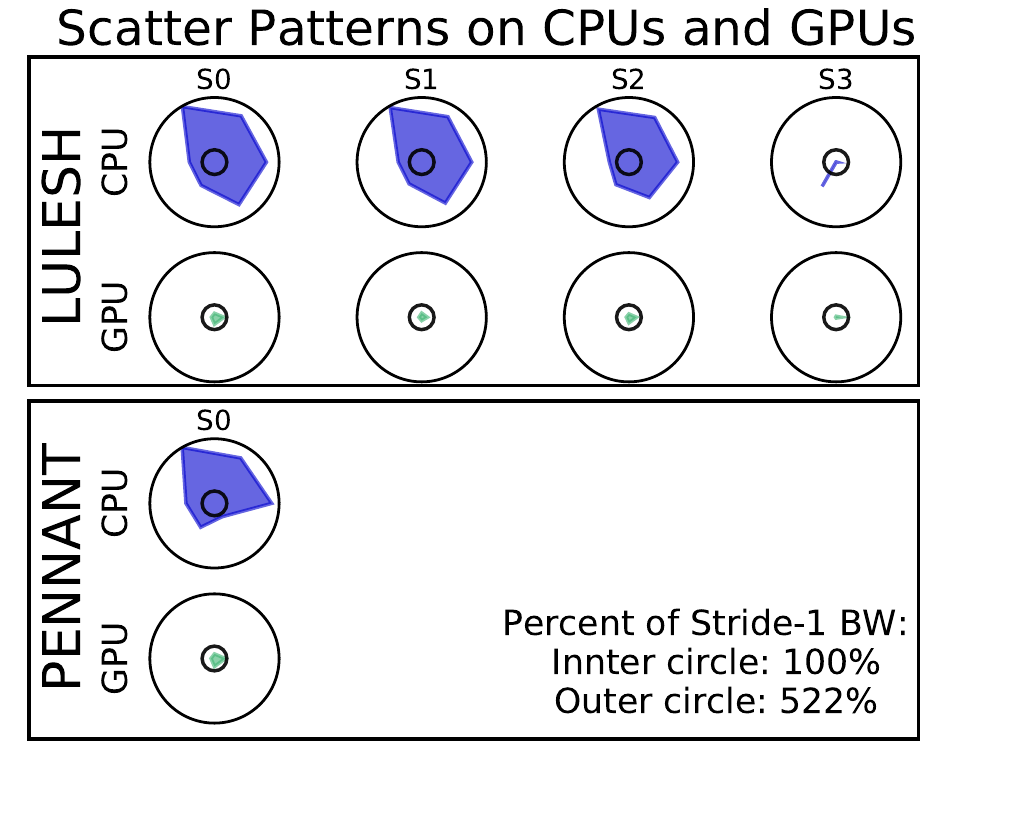}
\caption{App-derived Scatter Patterns}{Each circle represents the a single pattern. A spoke represents the performance of that pattern on a specific architecture as a percentage of the architecture's stride-1 bandwidth.}
\label{fig:radar_scatter}
\end{figure}

To get a handle on such questions, we take the per-platform and per-pattern data, and rendered them as using small-multiple radar plots as shown in \cref{fig:radar_gather} and \cref{fig:radar_scatter}. A single radar in this plot shows the performance of a pattern relative to its stride-1 performance across all CPUs (blue) and GPUs (green). The inner circle represents 100\% of stride-1 bandwidth, meaning that any value larger than this must be utilizing caching. This detailed look at the performance gives us a number of insights:

\begin{enumerate}
\item As promised in the last section, we first look at LULESH-S3, in \cref{fig:radar_scatter}. It indeed has very low performance, except on the TX2, which appears to handle the scenario of writing to the same location over and over very well. This could be due to an architectural optimization that recognizes data is being overwritten before it is ever used. 

\item Overall, we see that the GPUs are largely unable to outperform their stride-1 bandwidth. However, this appears to be changing with newer generations, as we see the V100 peaking above the 100\% circle for many of the patterns.

\item If we turn to the CPUs again, we see that Naples (in the bottom right of the CPU radars) largely under-performs, save for one set of patterns, Nekbone. Curiously, there is not much that differentiates the Nekbone patterns from LULESH, as both have uniform stride patterns with small deltas, so these patterns may require more thorough profiling and investigation. 

\item Restricting ourselves to the Intel processors, we see in the Gather patterns that improvements to the caching architecture have been made between Broadwell and the new Skylake and Cascade Lake architectures. We see a further improvement in Cascade Lake when looking at the LULESH scatter patterns, as it outperform Skylake as well. This indicates that even within the same architecture family, tweaks to caching and prefetching models can improve performance for hard-to-optimize scatter operations.

\item Finally, it is important to note the large dependence on the delta (distance between gather/scatter operations) that these patterns have. Looking at the PENNANT patterns, we see a large difference in performance starting at PENNANT-G5. If we look at \Cref{table:patterns}, we see that all the patterns before this have deltas less than or equal to 4, and the patterns including PENNANT-G5 and after have deltas larger than 400. \Cref{sec:bwbw} further expands on these patterns with a more detailed look at absolute and relative performance for these later PENNANT patterns. 
\end{enumerate}


\textit{Lesson: Spatter can be used to differentiate performance across architectures and can show how improved caching and prefetching hardware support in CPUs and limited caching in GPUs affects patterns with reuse. Additionally, these results can be correlated with the patterns themselves to show that delta is a primary indicator of performance for \gatherscatter operations.}

\subsubsection{Comprehensive Evaluation of Relative and Absolute Application Pattern Performance}
\label{sec:bwbw}

\begin{figure}[!htpb]
\centering
\begin{tabular}{c}
\includegraphics[width=1\linewidth]{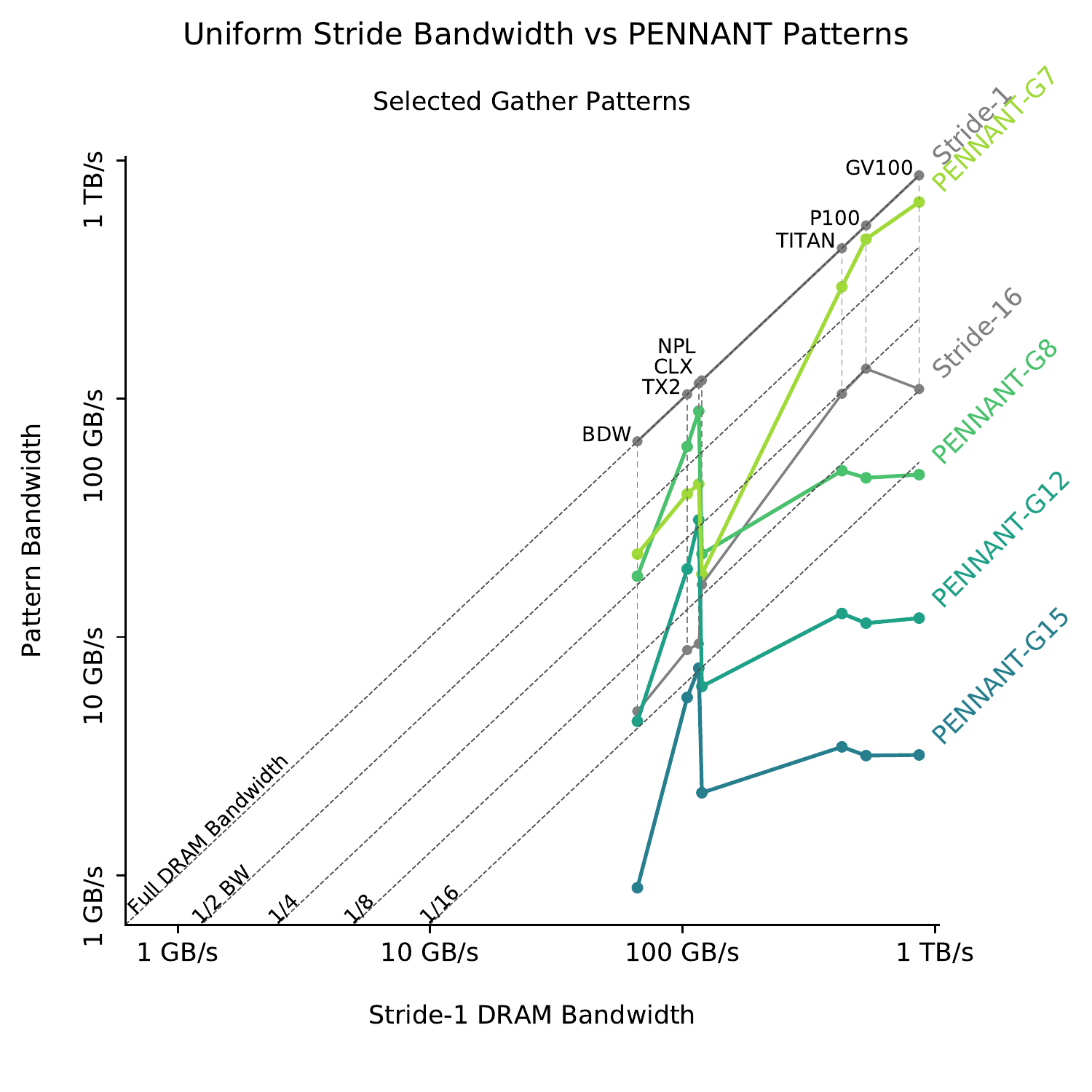} \\
\textbf{(a)} \\
\includegraphics[width=1\linewidth]{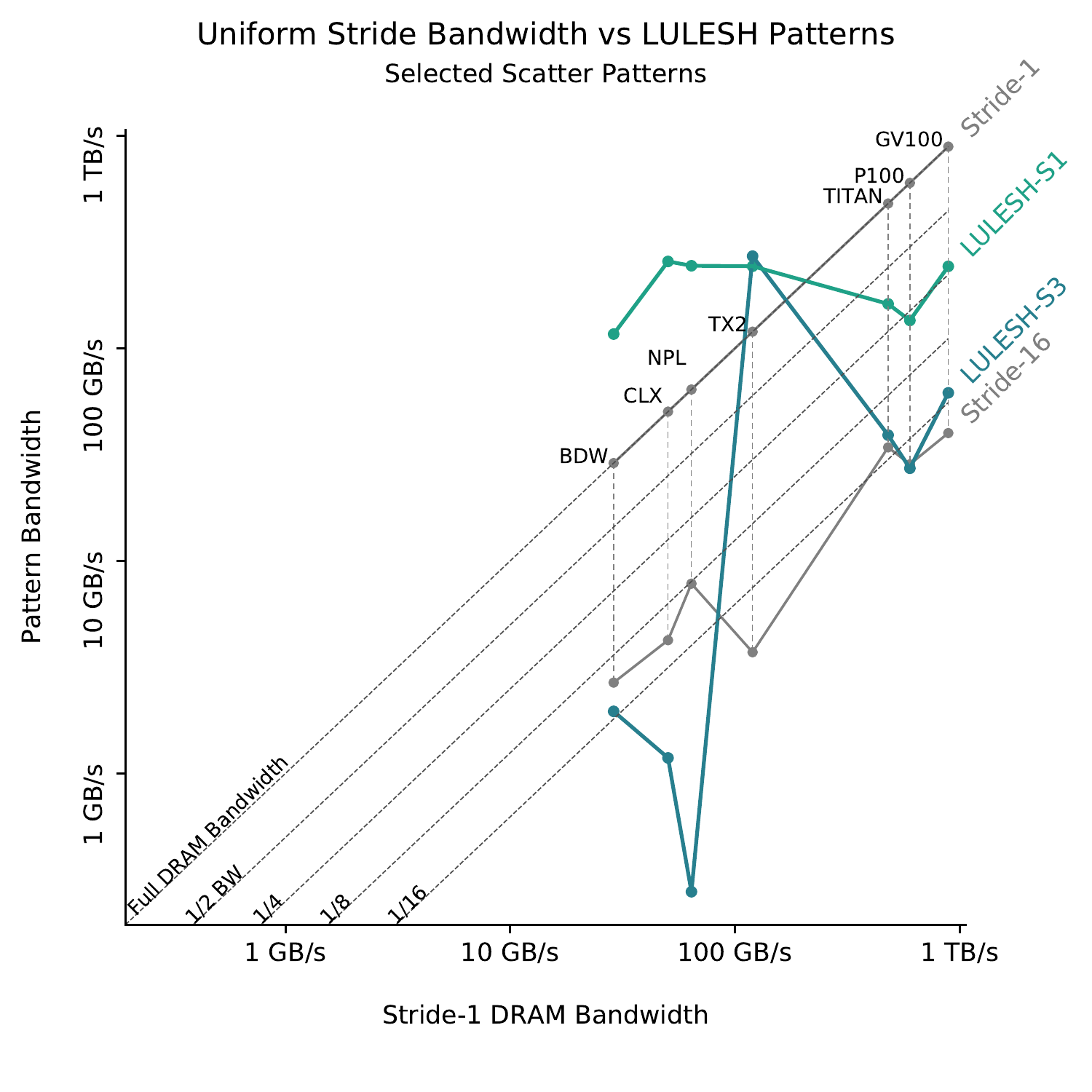} \\
\textbf{(b)} \\
\end{tabular}
\captionof{figure}{\small Bandwidth-Bandwidth Plots \textmd{\\We display a quantitative ranking of the selected platforms by plotting their pattern bandwidths as a function of the platform's stride-1 bandwidth. For a given platform, it stride-1 bandwidth is on the x=y diagonal, and selected pattern bandwidths appear directly below. Skykale is omitted from these plots as it is very similar to Cascade Lake.} \textit{Takeaway: For the patterns selected, CPUs show both an increase in performance, and relative performance across generations. Conversely, GPUs do not fare well on these patterns at all, leading to a decrease in the percentage of the bandwidth they use in the case of the gather patterns in (a).}}
\label{fig:bwbw}
\end{figure}

In addition to the high-level takeaways, we can also use Spatter measurements to plot both the the absolute and relative performance of patterns and how this relative performance varies between platforms. In \cref{fig:bwbw}, we have a few selected gather patterns from PENNANT in (a), and a few scatter patterns from LULESH in (b). 

These plots display application pattern performance as a function of stride-1 bandwidth. What this means is that stride-1 bandwidth will appear on the diagonal, and other bandwidths will appear in a vertical line through that point. It also means that all lines with unit slope are lines of constant fractional bandwidth. We have marked some of these lines in the plot for your reference. For instance, you can see that the PENNANT-G12 pattern runs at about 1/16th of the peak bandwidth on Broadwell. This complicated plot structure allows us to see both how well a pattern performs on platform X vs platform Y (by comparing the y values of the two points) and also how well a platform utilizes the bandwidth available to it on a given platform (by measuring a point's vertical distance from the diagonal). 

In \cref{fig:bwbw}(a), we have 4 different PENNANT patterns plotted, along with Stride-1 and Stride-16 results for reference. At a high level, there is clear left-to-right separation between CPUs and GPUs, due to the former having much less bandwidth available. There are a number of interesting points to discuss: 
\begin{enumerate}
\item If we take a look at just the Broadwell and Cascade Lake numbers, we see a slope that is greater than 1. This means is that Cascade Lake is not only better in absolute terms, but in relative terms as well, utilizing more of its available bandwidth than the Broadwell processor. 

\item A disappointing outlier is Naples, which performs much worse than its stride-1 bandwidth would suggest. This suggests a cache architecture much less capable than the other CPUs. We hope to compare this result with AMD EPYC processors in a future evaluation.

\item If we shift our attention to just the GPUs, we see that the large strides present in the higher-numbered PENNANT patterns have a large impact on the performance. If we reference \cref{table:patterns}, we see that the delta increases as the increase the pattern number increase. This shows us that while CPUs are able to handle these large deltas relatively well, GPUs have much worse relative performance as the delta increases.

\item Finally, if we look at both CPUs and GPUs, we see the power of this type of plot: we can see that the CPUs, due to the fact that the patterns contain some data reuse, are able to outperform GPUs on the selected patterns in terms of relative bandwidth.  
\end{enumerate}

\cref{fig:bwbw}(b) shows two LULESH scatter patterns. 
\begin{enumerate}

\item The only platform that does well on  LULESH-S3, which has a Scatter with delta 0. is the TX2, which we described in the previous section. 
\item LULESH-S1 gives us another interesting pattern that separates CPUs and GPUs with a uniform stride-24 pattern with delta 8. This means there is quite a bit of reuse between scatters that can be cached well by CPUs but that are handled poorly by the smaller caches on GPUs. 
\end{enumerate}

\textit{Lesson: By examining a number of application-derived \GS patterns, we show that (1) Spatter is able to reproduce unique behavior on CPUs that is not easily modeled by STREAM, (2) Spatter can also be used to discern improvements between architecture generations that go beyond simple bandwidth improvements, and (3) the Spatter benchmark suite can be used to quantitatively rank pattern performance between CPUs and GPUs and identify regimes where the CPUs are the clear winner in terms of relative performance.}

%

%% file: sections/related-work.tex
\section{Related Work}
\label{sec:related-work}
Our primary aim for Spatter is to measure at a low level the effects of sparsity and indirect accesses on effective bandwidth for a particular application or algorithm. While a number of bandwidth-related benchmarks exist, there are no current suites that explicitly support granular examinations of sparse memory accesses.
The closest analogue to our work is APEX-Map \cite{strohmaier:2005:apexmap}, which allows for varying sparsity to control the amount of spatial locality in the tested data set. However, APEX-map has not been updated for heterogeneous devices and does not allow for custom \GS patterns.

Similar to Spatter and APEX-Map, the HopScotch microbenchmark \cite{ahmed:2019:hopscotch} suite provides a tunable mechanism for representing mixes of read-only, write-only, and mixed access kernels in a similar fashion as Spatter. Currently, HopScotch includes a large suite of kernels intended to produce many different types of memory access patterns. While their suite does include \GS, we believe our work is complementary, allowing users a large degree of flexibility in the types of access patterns available, and the ability to run on GPUs. One technical difference of interest is that Spatter uses a slightly different interface to the kernel: instead of specifying the entire access pattern up front to the kernel, we specify an index pattern and a delta, so that our apps can more effectively mirror apps that generate indices at runtime, and not incur the overhead of moving a large index buffer through the memory hierarchy.

In terms of peak effective, or real-world achievable bandwidth, STREAM \cite{McCalpin1995} provides the most widely used measurement of sustained local memory bandwidth using a regular, linear data access pattern.
Similarly,  BabelStream \cite{deakin:2016:babelstream},  provides a STREAM interface for heterogeneous devices using  backends like OpenMP, CUDA, OpenCL, SyCL, Kokkos, and Raja. Intel's Parallel Research Kernels \cite{Hammond:2019:PRK} also supports an nstream benchmark that is used for some platforms here. The CORAL 2 benchmarks also include a STREAM variant called STRIDE \cite{seager:2019:stridebmk}, that includes eight different memory-intensive linear algebra kernels written in C and Fortran. STRIDE includes dot product and triad variations but still utilizes uniform stride inputs and outputs. None of these suites support any access pattern aside from uniform stride, which underlines the need for a benchmark like Spatter which includes \textit{configurable and indirect access patterns}.

Whereas STREAM focuses on a single access pattern, pointer chasing benchmarks \cite{hein:2018:emu_ashes18} and RandomAccess \cite{Luszczek05introductionto} use randomness in their patterns. Pointer chasing benchmarks measure the effects of memory latency but are limited in scope to measuring memory latency, and RandomAccess is only able to produce random streams. Spatter cannot model dependencies like pointer chasing but it contains kernels for modeling random access. 

\subsection{Heterogeneous Architectural Benchmarking}

Memory access patterns have been studied extensively on heterogeneous and distributed memory machines, where data movement has been a concern for a long time. Benchmarks such as SHOC \cite{Danalis:2010:SHC:1735688.1735702}, Parboil \cite{stratton2012parboil}, and Rodinia \cite{5306797} provide varying levels of memory access patterns that are critical to HPC applications.
For example, SHOC contains ``Level 0'' DeviceMemory and BusSpeedDownload benchmarks that can be used to characterize GPUs and some CPU-based devices. Likewise, other recent Department of Energy work has investigated vectorization support with hardware and compiler suites \cite{rajan2015vector_invest} for next-generation applications for the SIERRA supercomputer.
The design of Spatter is intended to create a new benchmark suite with a more focused set of access patterns to supplement these existing benchmark suites and studies and to provide a simpler mechanism for comparing scatter and gather operations across programming models and architectures.

Other work focuses on optimizing memory access patterns for tough-to-program heterogenous devices like GPUs. Recent work by Lai, et al. \cite{Lai:2018:sg_gpu_icpp} evaluates the effects of TLB caching on GPUs, develops an analytical model to predict the caching characteristics of \gatherscatter and then develops a multi-pass technique to improve the performance of \GS on modern GPU devices. Dymaxion \cite{Che:2011:dymaxion} takes an API approach to transforming data layouts and data structures and looks at scatter and gather as part of a sparse matrix-vector multiplication kernel experiment. Jang, et al. \cite{jang:2011:map_tpds} characterize loop body random and complex memory access patterns and attempt to resolve them into simpler and regular patterns that can be easily vectorized with GPU programming languages. Finally, CuMAPz \cite{Kim:2011:cumapz} provides a tool to evaluate different optimization techniques for CUDA programs with a specific focus on access patterns for shared and global memory.

\subsection{Extensions to Other Architectures}
One additional motivation for this work is to better implement sparse access patterns on nontraditional accelerators like FPGAs and the Emu Chick. For FPGAs, the Spector FPGA Suite \cite{spector} provides several features that have influenced the design of our benchmark suite including user-defined parameters for block size, work item size, and delta settings.

Spector uses OpenCL-based High-Level Synthesis and compiles a number of different FPGA kernels with various parameters and then attempts to pick the best configuration to execute on a specific FPGA device. While this process can be time-consuming for FPGAs due to routing heuristics, it does provide some motivation for the design of Spatter.
As shown in Section \ref{sec:bmk_design}, we design scripts that run multiple tests and then pick the best result for a given work item size, block size, and vector length to plot as the ``best'' result for a particular \gatherscatter operation. 

Finally, there is also work in computer architecture that explores the area of adding more capabilities to vector units. SuperStrider \cite{8123669} and Arm's Scalable Vector Extension \cite{7924233} both aim to implement \gatherscatter operations in hardware. Similarly, the Emu system \cite{dysart2016emu} focuses on improving random memory accesses by moving lightweight threads to the data in remote DRAM.
Spatter complements these hardware designs and associated benchmarking by allowing users to test how their code can benefit from dedicated data rearrangement units or data migration strategies. These projects primarily focus on architectural simulation and emulation, while we are looking at approaches to create effective sparse kernels that can be tested on FPGA prototypes or with these new architectures.

%% file: sections/future_work.tex
\section{Conclusions and Future Work} \label{sec:futurework}

This work has demonstrated the growing importance of indexed accesses in modern HPC applications and specifically looks at the use of gather and scatter operations in modern applications like the DoE mini-apps investigated in \cref{sec:app_char}. Spatter is introduced as a configurable benchmark suite that can be used to better evaluate these types of indirect memory accesses by using pattern-based inputs to generate a wide class of indexed memory access patterns. The presented Spatter experiments demonstrate how this tool could be used by architects to evaluate new prefetching hardware or instructions for gather and scatter, how compiler writers can inspect the performance implications of their generated code, and potentially how application developers could profile representative portions of their application that rely on \gatherscatterns. 

We envision that the Spatter benchmark will be a tool that can be used to examine any memory performance artifact that exists in sparse codes. The current model that Spatter implements, which is a single index buffer and delta for each pattern, is descriptive of a wide range of patterns that we have seen in DoE mini-apps as well as related benchmarks like STREAM and STRIDE. However, certain aspects of the memory hierarchy cannot be properly examined by the current version of Spatter, especially those relating to \textit{temporal locality}. 

To increase Spatter's ability to model memory access patterns, we plan to expand the benchmark suite with the following features: 
\begin{inparaenum}\item model temporal locality for accesses using time delta patterns to better represent cacheable access patterns, 
\item investigate mathematical and AI techniques for modeling more complex access patterns than can be represented with combinations of stride and delta parameters, and
\item develop new open-source techniques for extracting sparse memory access patterns from applications in a timely fashion. For this last goal, we are currently working on modeling 2D and 3D stencil operations from a proprietary full waveform inversion code used for ocean surveying.
\end{inparaenum} Other features that we are investigating for inclusion into Spatter are kernels written with intrinsics as well as new backends for Kokkos, SyCL, and novel architectures like FPGAs or the Emu Chick.

Our goal is also to make Spatter as easy to use as possible, and useful for a wide audience. To aid in this effort, we plan to make the following upgrades to the codebase:
\begin{inparaenum}
\item support for OpenMP 4.5 and SyCL backends, 
\item automation of parameter selection,
\item optimized CPU backends that make use of prefetching and streaming accesses, and
\item make as much of our tracing and trace analysis infrastructure available along with our codebase, which is open-source and available on Github.
\end{inparaenum}

Spatter is a powerful and flexible tool for investigating memory system architecture, and future research into improving this tool will further increase its ability to expose interesting memory patterns and architectural features. 


%% file: sections/appendix.tex
\section{Application Gather / Scatter Patterns}
\label{sec:pattern_appendix}
\Cref{table:patterns} lists all the patterns used in evaluation of the Spatter suite. 

\begin{table*}[htp]
\centering
\caption{Listing of Patterns}
\label{table:patterns}
\begin{small}
\begin{tabular}{l|ccc}
\toprule
   \textbf{Gather Pattern} & Index & Delta & Type\\
\midrule
\midrule
PENNANT-G0 & [2,484,482,0,4,486,484,2,6,488,486,4,8,490,488,6] & 2 & \\
PENNANT-G1 & [0,2,484,482,2,4,486,484,4,6,488,486,6,8,490,488] & 2 & \\
PENNANT-G2 & [0,4,8,12,16,20,24,28,32,36,40,44,48,52,56,60] & 2 & Stride-4\\
PENNANT-G3 & [4,8,12,0,20,24,28,16,36,40,44,32,52,56,60,48] & 2 & \\
PENNANT-G4 & [0,0,0,0,1,1,1,1,2,2,2,2,3,3,3,3] & 4 & Broadcast\\
PENNANT-G5 & [4,8,12,0,20,24,28,16,36,40,44,32,52,56,60,48] & 4 & \\
PENNANT-G6 & [482,0,2,484,484,2,4,486,486,4,6,488,488,6,8,490] & 480 & \\
PENNANT-G7 & [482,0,2,484,484,2,4,486,486,4,6,488,488,6,8,490] & 482 & \\
PENNANT-G8 & [2,0,0,0,2,0,0,0,2,0,0,0,2,0,0,0] & 129608 & \\
PENNANT-G9 & [0,0,0,0,1,1,1,1,2,2,2,2,3,3,3,3] & 388852 & Broadcast\\
PENNANT-G10 & [0,0,0,0,1,1,1,1,2,2,2,2,3,3,3,3] & 388848 & Broadcast\\
PENNANT-G11 & [0,0,0,0,1,1,1,1,2,2,2,2,3,3,3,3] & 388848 & Broadcast\\
PENNANT-G12 & [6,0,2,4,14,8,10,12,22,16,18,20,30,24,26,28] & 518408 & \\
PENNANT-G13 & [6,0,2,4,14,8,10,12,22,16,18,20,30,24,26,28] & 518408 & \\
PENNANT-G14 & [6,0,2,4,14,8,10,12,22,16,18,20,30,24,26,28] & 1036816 & \\
PENNANT-G15 & [0,0,0,0,1,1,1,1,2,2,2,2,3,3,3,3] & 1882384 & Broadcast\\

   \midrule
LULESH-G0 & [0,1,2,3,4,5,6,7,8,9,10,11,12,13,14,15] & 1 & Stride-1\\
LULESH-G1 & [0,1,2,3,4,5,6,7,8,9,10,11,12,13,14,15] & 8 & Stride-1\\
LULESH-G2 & [0,8,16,24,32,40,48,56,64,72,80,88,96,104,112,120] & 1 & Stride-8\\
LULESH-G3 & [0,24,48,72,96,120,144,168,192,216,240,264,288,312,336,360] & 8 & Stride-24\\
LULESH-G4 & [0,24,48,72,96,120,144,168,192,216,240,264,288,312,336,360] & 4 & Stride-24\\
LULESH-G5 & [0,24,48,72,96,120,144,168,192,216,240,264,288,312,336,360] & 1 & Stride-24\\
LULESH-G6 & [0,24,48,72,96,120,144,168,192,216,240,264,288,312,336,360] & 8 & Stride-24\\
LULESH-G7 & [0,1,2,3,4,5,6,7,8,9,10,11,12,13,14,15] & 41 & Stride-1\\

   \midrule
NEKBONE-G0 & [0,6,12,18,24,30,36,42,48,54,60,66,72,78,84,90] & 3 & Stride-6\\
NEKBONE-G1 & [0,6,12,18,24,30,36,42,48,54,60,66,72,78,84,90] & 8 & Stride-6\\
NEKBONE-G2 & [0,6,12,18,24,30,36,42,48,54,60,66,72,78,84,90] & 8 & Stride-6\\
   \midrule
AMG-G0 & [1333,0,1,36,37,72,73,1296,1297,1332,1368,1369,2592,2593,2628,2629] & 1 & Mostly Stride-1\\
AMG-G1 & [1333,0,1,2,36,37,38,72,73,74,1296,1297,1298,1332,1334,1368] & 1 & Mostly Stride-1\\
 \toprule
   \textbf{Scatter Pattern} & Index & Delta & Type\\
\midrule
\midrule
PENNANT-S0 & [0,4,8,12,16,20,24,28,32,36,40,44,48,52,56,60] & 1 & Stride-4\\
    \midrule
LULESH-S0 & [0,8,16,24,32,40,48,56,64,72,80,88,96,104,112,120] & 1 & Stride-8\\
LULESH-S1 & [0,24,48,72,96,120,144,168,192,216,240,264,288,312,336,360] & 8 & Stride-24\\
LULESH-S2 & [0,24,48,72,96,120,144,168,192,216,240,264,288,312,336,360] & 1 & Stride-24\\
LULESH-S3 & [0,24,48,72,96,120,144,168,192,216,240,264,288,312,336,360] & 0 & Stride-24\\

\bottomrule
\end{tabular}
\end{small}
\end{table*}